\documentclass[prd,amsmath,amssymb,superscriptaddress,nofootinbib]{revtex4}
\usepackage{amsfonts}
\usepackage{multirow}
\usepackage{mathrsfs}
\usepackage{graphicx}
\usepackage{amsmath}
\usepackage{amssymb}
\usepackage{bm}
\usepackage{color}
\usepackage{diagbox}
\usepackage{hyperref}
\usepackage{slashed}

\usepackage{ulem} 

\newcommand{\nn}{\nonumber}

\newcommand{\beq}{\begin{equation}}
\newcommand{\eeq}{\end{equation}}
\newcommand{\bqa}{\begin{eqnarray}}
\newcommand{\eqa}{\end{eqnarray}}

\newcommand{\bseq}{\begin{subequations}}
\newcommand{\eseq}{\end{subequations}}

\makeatletter

\begin{document}

\title{A tale of Bethe logarithms: leptonic widths of $\chi_{cJ}$ and Lamb shift}
\author{Yu Jia~\footnote{jiay@ihep.ac.cn}}
\affiliation{Institute of High Energy Physics, Chinese Academy of Sciences, Beijing 100049, China\vspace{0.2cm}}
\affiliation{School of Physical Sciences,
University of Chinese Academy of Sciences, Beijing 100049, China\vspace{0.2cm}}
\author{Jichen Pan~\footnote{panjichen@pku.edu.cn}}
\affiliation{Institute of High Energy Physics, Chinese Academy of Sciences, Beijing 100049, China\vspace{0.2cm}}
\affiliation{School of Physics, Peking University, Beijing 100871, China\vspace{0.2 cm} }
\affiliation{Center for High Energy Physics, Peking University, Beijing 100871, China\vspace{0.2 cm} }

\date{\today}

\begin{abstract}
The rare annihilation decays of $P$-wave spin-triplet quarkonia into lepton pair
have to proceed via two-photon intermediate state, which are plagued with the infrared divergence symptom.
We recognize that the physical root of the IR divergence and its remedy is the same as the Lamb shift in QED.
In this work we provide a complete solution to this  IR problem by including the effect of the higher Fock component of the $\chi_{cJ}$ state,
{\it viz.}, the $c\bar{c}(^3S_1^{(1)})$ pair accompanied with a very long wavelength photon.
Adding the contributions from the leading and next-to-leading order Fock components together,
we arrive at the IR finite and factorization scale independent predictions for leptonic widths of $\chi_{cJ}(nP)$.
The Bethe logarithms associated with these exclusive reactions are found to have rather different traits
from those associated with Lamb shift. We present numerical predictions by employing several influential quark potential models.
The predicted leptonic widths of $\chi_{cJ}(nP)$ are sizable and the observation prospect for
$e^+e^-\to \chi_{cJ}(1P,2P)$ at {\tt BESIII} looks bright. The future observation of $e^+e^-\to X(3872)$
will shed important light on the $c\bar{c}$ content of the $X(3872)$ meson.
\end{abstract}

\maketitle

\section{Introduction}


Leptonic decays of vector quarkonia, {\it e.g.}, $J/\psi(\Upsilon)\to \gamma^*\to e^+e^-$, are among the simplest and best-studied
processes in hadron physics. What is less familiar, the $C$-even quarkonia can also couple to a lepton pair,
albeit through two photon intermediate state. The leptonic decays $\chi_{c0}/\eta_c\to e^+e^-$
are strongly suppressed due to the penalty brought by the QED helicity conservation~\cite{Kuhn:1979bb,Jia:2009ip}.
Recently {\tt BESIII} experiment reported the first observation of exclusive $\chi_{c1}$ production in $e^+e^-$ collision with a significance of $5.1\;\sigma$,
from which the leptonic decay width of $\chi_{c1}$ is determined to be $\Gamma(\chi_{c1}\to e^+ e^-)=0.12^{+0.13}_{-0.08}$ eV~\cite{BESIII:2022mtl}.
{\tt BESIII} experiment plans to further search the $e^+e^-\to \chi_{c2}$, $\chi_{c1}(3872)$ channels through energy scan strategy.
To provide a trustworthy theoretical guidance, it is mandatory to develop a thorough understanding of the mechanism underlying these
rare exclusive reactions.

In 1979 Kuhn, Kaplan and Safiani~\cite{Kuhn:1979bb} discovered that the $\chi_{c1,2}\to e^+e^-$ processes are plagued with the
logarithmic infrared (IR) divergences within the framework of the color-singlet model,
the precursor of nonrelativistic QCD (NRQCD) factorization approach ~\cite{Bodwin:1994jh}.
The breakdown of NRQCD factorization in such processes clearly indicates
that some important long-distance dynamics is missing.

Some theoretical efforts were exerted to alleviate the IR pathology for $\chi_{c1,2}\to e^+e^-$
during the past decade~\cite{Yang:2012gk,Kivel:2015iea}.
Using the threshold expansion technique~\cite{Beneke:1997zp}, Yang and Zhao~\cite{Yang:2012gk} found that the
{\it ultrasoft} region in the two-photon box diagrams  develop the logarithmic UV pole,
which exactly cancels the IR pole affiliated with the hard region contribution.
Later Kivel and Vanderhaeghen~\cite{Kivel:2015iea} confirmed this threshold expansion analysis,
and further pointed out that including the ultrasoft mode amounts to including the contribution
from the higher Fock components of $\chi_{cJ}$.
In analogy to the color octet mechanism tailed to eliminate the IR divergence
in $P$-wave quarkonium {\it inclusive} decays~\cite{Bodwin:1992ye,Huang:1996cs},
some nonpertubative ultrasoft matrix elements are introduced in \cite{Kivel:2015iea},
whose magnitudes are estimated using the heavy hadron chiral perturbation theory (HH$\chi$PT).

The solution offered in Ref.~\cite{Kivel:2015iea}, notwithstanding capturing some essential physics,
is still unsatisfactory in several aspects. Firstly,
since quarkonia are chiral singlets, the HH$\chi$PT approach employed in \cite{Kivel:2015iea} appears to
lack strong predictive power. In fact, various inputs parameters in \cite{Kivel:2015iea} have to be taken from either experiments or potential quark models.
Secondly, the predictions made in \cite{Kivel:2015iea} depend on some artificial factorization scale,
and no attempt is made to estimate theoretical uncertainties.
Thirdly, the role played by the {\it ultrasoft} photon in the $P$-wave quarkonium
{\it exclusive} electromagnetic decay, notably differs from
that played by the {\it soft} gluon in the $P$-wave onium {\it inclusive} hadronic decay~\cite{Bodwin:1992ye,Huang:1996cs},
hence the analogue between these two situations looks obscure.
Lastly, extending the approach of \cite{Kivel:2015iea} to infer leptonic widths of the
excited $\chi_{cJ}$ states appears not straightforward.

This work strives to achieve a systematic accoiunt of $\chi_{cJ}\to e^+e^-$ within the confine of NRQCD,
{\it viz.}, to present the {\it IR-finite} and {\it factorization scale independent} predictions
by including the contributions of the next-to-leading Fock component of $\chi_{cJ}$.
Theoretical errors are estimated by varying potential models.
The leptonic widths of $\chi_{cJ}(2P)$ are also predicted. Our numerical studies indicate that
future observation or nonobservation of $e^+e^-\to X(3872)$ can effectively
pin down the $c\bar{c}$ content of the enigmatic $X(3872)$
meson~\cite{Belle:2003nnu, CDF:2006ocq, LHCb:2013kgk}.

It is illuminating to recognize that the IR symptom in $\chi_{cJ}\to e^+e^-$
bears the same root as that in the Lamb shift~\cite{Lamb:1947zz}.
As is well known, Lamb shift of hydrogen atom can be split into (relativistic) high-energy part and (non-relativistic)
low energy part~\cite{Weinberg:QFT}.
While the former entails logarithmic IR divergence, the latter contains logarithmic UV divergence.
Upon summing these two parts, one ends up with finite and factorization scale independent predictions
for the level shift of hydrogen atom~\cite{Weinberg:QFT}, in the disguise of Bethe logarithms~\cite{Bethe:1947id}.

As will be elaborated later, the exactly same rhyme repeats itself for $\chi_{cJ}\to e^+e^-$.
The leading order (LO) NRQCD short-distance coefficient, which corresponds to
the hard region contribution in the two-photon box diagram, is the counterpart of the high energy part in Lamb shift.
What we need to include is the counterpart of the low energy part in Lamb shift.
The low energy part in Lamb shift can be deduced from the self energy of a bound non-relativistic electron
through emitting and reabsorbing a very long wave-length (ultrasoft) photon via electric dipole (E1) transition.
Equivalently, as pioneered by Bethe in his seminal 1947 paper~\cite{Bethe:1947id},
the low energy part of the Lamb shift can also be viewed as the energy shift due to the higher Fock components of a physical electron,
{\it viz.}, a bound electron plus an ultrasoft photon.
Therefore, it looks appealing that one should also include the contribution from the higher Fock component of $\chi_{cJ}$,
{\it viz.}, a $c\bar{c}({}^3S_1)$ pair accompanied with an ultrasoft photon.
Exactly like what happens in Lamb shift, after the contributions from both the LO and next-to-leading order (NLO) Fock states
are included, the unwanted IR poles as well as the factorization scale will disappear from
the final amplitudes of $\chi_{c1,2}\to e^+e^-$.

When the QED multipole interaction is turned on, a physical $\chi_{cJ}$ state can be expanded into
\beq
\vert \chi_{cJ}\rangle =  \left\vert c\bar{c}({}^3P_J^{(1)}) \right\rangle +  \left\vert c\bar{c}({}^3S_1^{(1)})+\gamma_{\rm us}\right\rangle +\cdots,
\label{Fock:expansion:chicJ}
\eeq

It turns out that the contributions from the LO and NLO Fock components of $\chi_{cJ}$
scale with the same powers in charm quark velocity. The amplitude can be split into two pieces:
\beq
{\cal A}[\chi_{cJ}\to e^+e^-] = {\cal A}^{(J)}_{\rm LF} + {\cal A}^{(J)}_{\rm HF}.
\label{Splitting:Amplitudes:to:two:parts}
\eeq
The full decay amplitudes will be expressed in terms of the Bethe logarithms.
Interestingly, the Bethe logarithms in $\chi_{cJ}\to e^+e^-$ turn out to
bear quite different characteristics from those in Lamb shift.

\section{Contribution from Leading Fock state}

\begin{figure}[hbt]
\begin{center}
\includegraphics[width=0.8\textwidth]{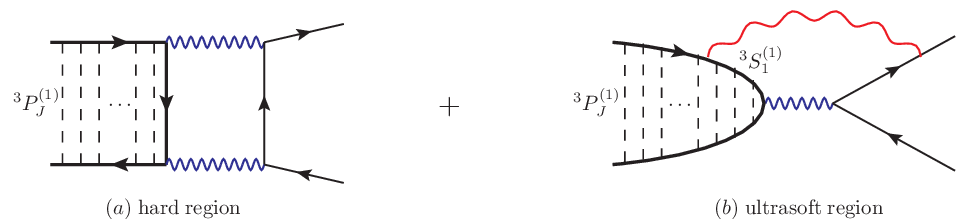}
\caption{Portrait of two leading contributions to $\chi_{cJ}\to e^+e^-$.
Diagram $(a)$ represents the contribution from the leading Fock component of $\chi_{cJ}$, $\vert c\bar{c}({}^3P^{(1)}_J)\rangle$,
where two virtual photons  simultaneously carry hard momenta;
Diagram $(b)$ encapsulates the contribution induced by the next-to-leading Fock component of $\chi_{cJ}$, {\it viz.},
$\vert c\bar{c}({}^3S^{(1)}_1)+\gamma^{\rm us} \rangle$, where one photon is hard and another is ultrasoft,
represented by the blue and red wavy curves, respectively.
The vertical dashed line represents the exchange of the color-singlet static potential $V_S(r)$ between $c$ and $\bar{c}$.
\label{Feyn:diagrams:chicJ:to:ee}}
\end{center}
\end{figure}

Let us specialize to the rare decay channels  $\chi_{c1,2}(nP) \to e^-(l_1)e^+(l_2)$.
We work in the rest frame of $\chi_{cJ}(nP)$ with $P^\mu=(M_{\chi_c},{\bf 0})$,
and assume $l_1^2=l_2^2=0$ for simplicity.
The leptonic decays initiated by the leading Fock components of $\chi_{cJ}$ are
illustrated by Fig.~\ref{Feyn:diagrams:chicJ:to:ee}(a), which proceed through the quark processes
$c\bar{c}({}^3P^{(1)}_J)\to \gamma^*\gamma^*\to e^+e^-$ with two hard photon exchange.
The corresponding reduced amplitudes are computed in the context of NRQCD factorization~\cite{Yang:2012gk}~\footnote{To condense the expressions,
we have introduced the reduced amplitudes $\widetilde{\cal A}^{(J)}$ via ${\cal A}^{(1)} = \bar{u}(l_1) \gamma^{i}v(l_2) \epsilon^{ijk} (l_1-l_2)^j
\varepsilon^k_{\chi_{c1}} \widetilde{\cal A}^{(1)}$, and  ${\cal A}^{(2)} = \sqrt{2} \bar{u}(l_1) \gamma^{i}v(l_2)
h^{ij}_{\chi_{c2}} (l_1-l_2)^j
\widetilde{\cal A}^{(2)}$, where $\varepsilon^i_{\chi_{c1}}$ and $h^{ij}_{\chi_{c2}}$ represent the polarization vector/tesnor of $\chi_{c1}$ and $\chi_{c2}$, respectively.}:
\begin{subequations}
\bqa
& & \widetilde{\cal A}^{(1)}_{\rm LF} = -i\frac{ e_c^2 \alpha^2 \sqrt{N_c}}{m_c^4} \sqrt{\frac{3}{4\pi}} R'_{nP}(0)\left({1\over \hat\epsilon_{\rm IR}}+2\ln{\mu\over m_c}\right),
\\
& & \widetilde{\cal A}^{(2)}_{\rm LF} = -i\frac{ e_c^2 \alpha^2 \sqrt{N_c}}{m_c^4} \sqrt{\frac{3}{4\pi}} R'_{nP}(0)
\left[{1\over \hat\epsilon_{\rm IR}} + 2\ln{\mu\over m_c}-\frac{2(1-\ln 2 -i \pi)}{3}\right],
\eqa
\label{Leading:Fock:state:hard:contr:to:amplitude:chicJ:ee}
\end{subequations}
where $N_c=3$ denotes the number of color, $e_c e= {2\over 3}e$ denotes the electric charge of the charm quark.
$R'_{\chi_c}(0)$ denotes the first derivative of radial wave functions at the origin for $\chi_{cJ}$.
Note the dimensional regularization (DR) is employed to regularize the UV and IR divergences, with
${1\over \hat\epsilon} \equiv {1\over \epsilon}- \gamma_E+\ln 4\pi $, and $\mu$ specifies the 't Hooft unit of mass.
The occurrence of the imaginary part in the amplitude of $\chi_{c2}\to e^+e^-$ arises from the two-photon cut,
while the absence of the imaginary part in leptonic decay of $\chi_{c1}$ is attributed to the prohibition of
the reaction $\chi_{c1}\to \gamma \gamma$ by Landau-Yang theorem.

\section{Contribution from higher Fock state}

As depicted in Fig.~\ref{Feyn:diagrams:chicJ:to:ee}(b), another leading region in the box diagram
is when one photon is ultrasoft while the other remains hard, which can be interpreted as the
the contribution from the NLO Fock component of $\chi_{cJ}$ in \eqref{Fock:expansion:chicJ}.
The intermediate $\psi(mS)$ states in the high Fock component of $\chi_{cJ}$,
which are induced from the leading $c\bar{c}({}^3P_J)$ component
through $E1$ transition, annihilate via a hard virtual photon into a pair of energetic back-to-back leptons,
on one of which the ultrasoft photon ends.
It is important to emphasize that, since there is no suppression of $v$ when adding a potential gluon exchange between slowly-moving
$c$ and $\bar{c}$ in the color-singlet channel,  one should resum all the gluonic ladder diagrams
before and after the ultrasoft photon is emitted,
which encapsulates the nonperturbative binding effect inherent to charmonium.

Practically, it is convenient to invoke the quantum mechanical perturbation theory to ascertain the higher Fock state contribution.
Analogous to the textbook treatment of the atomic spontaneous radiation,
we treat the $c$ and $\bar{c}$ within the first-quantized nonrelativistic quantum mechanics,
yet treat the ultrasfot electromagnetic field in the second quantized manner~\footnote{As will be shown in
Appendix~\ref{Rederivation:from:pNRQCD+SCET}, the fully-fledged field-theoretical treatment using
potential NRQCD (pNRQCD) plus soft-collinear effective theory (SCET) would yield identical answer as this simple and intuitive derivation.}.
We break the full Hamiltonian into two parts, $H=H_0+V_{\rm E1}$. The ``free" part
$H_0= {{\bf p}^2 \over m_c}+ V_S(r) + \int\!d^3{x} {1\over 2}\left({\bf E}^2+{\bf B}^2 \right)$,
with $V_{S}(r)$ signifying the static potential in the color-singlet channel (To be phenomenologically viable,
it is beneficial to include the linearly confining potential in the long distance)~\footnote{We note that a preceding work~\cite{Beneke:2008pi}
on exclusive decay $B\to \chi_{cJ} K$ also invoked the higher Fock state $\vert c\bar{c}({}^3S_1^{(8)})+g_{\rm us}\rangle$
to tame the IR divergence encountered in the color-singlet channel calculation~\cite{Song:2003yc,Meng:2005er}
, where the analysis was conducted in the academic Coulomb limit $mv^2\gg \Lambda_{\rm QCD}$. In contrast, since the process considered here is only sensitive to the higher Fock state mediated by QED,
our approach is still justified in the more realistic situation $mv^2\sim \Lambda_{\rm QCD}$.}.
$V_{\rm E1}= {e e_c\over m_c }{\bf p}\cdot {\bf A}^{\rm em}$
mediates the $E1$ transition. Treating $V_{\rm E1}$ as a first-order perturbation,  the NLO Fock component of $\chi_{cJ}$ is expressed as
\beq
\left\vert c\bar{c}({}^3S_1^{(1)})+\gamma_{\rm us}\right\rangle = {1\over E_{\chi_{cJ}}-H_0+i\epsilon} V_{\rm E1} \left\vert c\bar{c}({}^3P_J^{(1)}) \right\rangle.
\label{Next:to:leading:Fock:component:chicJ}
\eeq

Inserting a complete set of energy eigenstates of $H_0$, {\it viz.}, vector charmonium plus a photon,
we obtain the nonvanishing amplitudes for $\chi_{cJ}\to e^-e^+$ from the NLO Fock component:
\beq
{\cal A}^{(J)}_{\rm HF}  =  \int_{{\bf k}\ll m_c v} \!\! {d^3 k\over (2\pi)^3 2 |{\bf k}|}  \sum_{\rm Pol}
\sum_{m=1}^\infty {  {\cal A}^{\rm E1}\left[ \chi_{cJ}(nP) \to \psi(mS) \gamma({\bf k}) \right] \,
{\cal A}\left[\psi(m S) \gamma({\bf k}) \to e^+ e^-\right]
\over E_{\chi_{cJ}}-E_{mS}-|{\bf k}| +i \epsilon}.
\label{Higher:Fock:state:contr:to:S:matrix:chicJ:ee}
\eeq
Note the recoil of the vector charmonia in $E1$ transition is neglected.
We also neglect the contributions from the intermediate ${}^3D_1^{(1)}$ levels
due to strong velocity suppression~\footnote{It is worth mentioning one remarkable difference between this case and the Lamb shift.
Due to color confinement characteristic of QCD, here we only consider the {\it discrete} $\psi(mS)$ levels
in the sum over intermediate states (for simplicity we neglect the string breaking effect).
In sharp contrast with the Lamb shift where the bulk of contribution
comes from the continuum states, the dominant contribution in our case arises from the sum of first few energy levels.}.

Equation~\eqref{Higher:Fock:state:contr:to:S:matrix:chicJ:ee} entails two basic building blocks.
One is the familiar $E1$ transition amplitudes between $S$ and $P$-wave quarkonia,
which have also been reproduced from the angle of pNRQCD~\cite{Brambilla:2005zw,Brambilla:2012be}:
\begin{subequations}
\begin{align}
{\cal A}^{\rm E1}\left[\chi_{c1}(nP) \to \psi(mS) + \gamma({\bf k}) \right] &=
- \frac{e e_c \sqrt{3}}{\sqrt{2}m_c}\epsilon^{ijk} \varepsilon_{\psi}^{*i}  \varepsilon_\gamma^{*j}(k) \varepsilon^k_{\chi_{c1}} {\cal Z}_{nm},
\label{S:matrix:E1:transition:chic1}
\\
{\cal A}^{\rm E1}\left[\chi_{c2}(nP) \to \psi(mS) + \gamma({\bf k}) \right] &=
- \frac{e e_c \sqrt{3}}{m_c} \varepsilon_{\psi}^{*i} h^{ij}_{\chi_{c2}} \varepsilon_\gamma^{*j}(k) {\cal Z}_{nm}.
\end{align}
\label{S:matrix:E1:transition}
\end{subequations}
with the overlap integral given by
\beq
{\cal Z}_{nm}  =   \int_0^\infty \!\! dr\, r^2 R_{mS} (r) \left(\frac{1}{3}R_{nP}'(r)+\frac{2}{3}\frac{R_{nP}(r)}{r}\right).
\label{E1:overlap:integral}
\eeq

The second building block is the amplitude of $\psi(mS)+ \gamma_{\rm us} \to e^+e^-$.
Since the virtuality of the hard photon into which $c$ and $\bar{c}$ annihilate
is $4m_c^2$,  one can also employ the NRQCD factorization  to tackle this subprocess.
One obtains
\beq
{\cal A}[\psi(mS)+ \gamma_{\rm us} \to e^+e^-] =
    i\frac{\sqrt{2} e^3 e_c}{2\sqrt{4\pi} m_c^2} R_{nS}(0)
    \left[ {l_1\cdot \varepsilon_\gamma(k) \over l_1 \cdot k}  -   {l_2\cdot \varepsilon_\gamma(k) \over l_2
    \cdot k}\right]\, \bar{u}(l_1) \bm{\gamma} \cdot \bm{\varepsilon}_{\psi}  v(l_2).
\label{S:matrix:3S1+photon:to:lepton:pair}
\eeq
The ultrasoft photon can be attached to either the outgoing $e^-$ or $e^+$, and one utilizes
the eikonal approximation to the lepton propagators.

It is a good place to pause and scrutinize the powers of $v$ in ${\cal A}^{(J)}_{\rm HF}$.
The integration measure, energy denominator, $E1$ transition amplitude, and the amplitudes of $\psi(mS)+ \gamma_{\rm us} \to e^+e^-$,
count as $v^{6-2}$, $v^{-2}$, $v^{1}$, and $v^{-2}$, respectively.
So ${\cal A}^J_{\rm HF}$ in \eqref{Higher:Fock:state:contr:to:S:matrix:chicJ:ee} scales as $v^{6-2-2+1-2}=v$,
the same as the amplitude from the leading $P$-wave Fock component, \eqref{Leading:Fock:state:hard:contr:to:amplitude:chicJ:ee}.
Intriguingly, the $v^{-2}$ enhancement factor brought by the QED eikonal propagator plays a vital role
in promoting the higher Fock state contribution to the same significance as the leading Fock state contribution.

Plugging \eqref{S:matrix:E1:transition} and \eqref{S:matrix:3S1+photon:to:lepton:pair} into \eqref{Higher:Fock:state:contr:to:S:matrix:chicJ:ee}, summing over
the polarizations of the $c\bar{c}(n{}^3S_1^{(1)})$ and the ultrasfot photon, one obtains
\begin{subequations}
\begin{align}
 {\cal A}^{(1)}_{\rm HF}
 &= -i \frac{e^4 e_c^2\sqrt{N_c}}{2 m_c^3}\sqrt{\frac{3}{4\pi}}\sum_{m=1}^\infty
 R_{mS}(0)  {\cal Z}_{nm}
\nn\\
& \times  \bar{u}(l_1) \gamma^{i}v(l_2)  \epsilon^{ijk} \varepsilon_{\chi_{c1}}^k  \Big[ I^j \left(E_{nP}-E_{mS}, \hat{l}_1\right)
-
I^j \left(E_{nP}-E_{mS}, \hat{l}_2\right)\Big],
\label{chic1:ee:HF:contribution:key:formula}
\\
{\cal A}^{(2)}_{\rm HF}
 &= -i \frac{\sqrt{2}e^4 e_c^2\sqrt{N_c}}{2 m_c^3}\sqrt{\frac{3}{4\pi}}\sum_{m=1}^\infty
 R_{mS}(0) {\cal Z}_{nm}
\nn\\
& \times  \bar{u}(l_1) \gamma^i v(l_2) \,h^{ij} \Big[ I^j \left(E_{nP}-E_{mS}, \hat{l}_1\right)
-
I^j \left(E_{nP}-E_{mS}, \hat{l}_2\right)\Big],
\label{chic2:ee:HF:contribution:key:formula}
\end{align}
\label{chicJ:ee:HF:contribution:key:formula}
\end{subequations}
where the integral $I^i$ is defined in the $D-1$ spatial dimensions to regularize divergences:
\bqa
 I^i(\Delta E,\hat{l}_a) & \equiv & \mu^{2\epsilon} \int\!\! {d^{D-1} k \over (2 \pi)^{D-1} }\,{1\over 2 |{\bf k}|}\cdot
{\delta^{ij}-\hat{k}^i\hat{k}^j \over \Delta E - |{\bf k}|+i \epsilon} \cdot {l_a^j\over k\cdot l_a} \qquad\quad(a=1,2)
\nn\\
& =& - \hat{l}^i_a {(4\pi)^\epsilon \over 8\sqrt{\pi}\,\sin(2\pi\epsilon)\,\Gamma({3\over 2}-\epsilon)}
\left({\mu \over -\Delta E-i\epsilon}\right)^{2\epsilon}
\nn\\
& =& - {\hat{l}^i_a \over 8\pi^2} \left[ {1\over \hat\epsilon_{\rm UV}} + 2 +
2\ln {\mu \over 2(-\Delta E-i\epsilon)}+{\cal O}(\epsilon)\right].
\label{One:loop:integral:dim:reg}
\eqa
With the potential IR divergence cutoff by the quarkonium binding energy, this ultrasoft integral turns out to be
logarithmically UV divergent~\!\footnote{Intriguingly, the transverse photon polarization sum in the numerator turns to be crucial to
sweep the potential collinear divergence as $k\parallel l_a$.}.

Employing a simple identity~\!\!\!\footnote{This identity can be proved by invoking the quantum mechanical completeness relation
$\delta^{(3)}({\bf r})=\langle 0 \vert {\bf r}\rangle = {1\over 4\pi} \sum_{m} R_{mS}(r) R_{mS}(0)$, which
follows from the fact that all the partial waves higher than the $S$ wave yield vanishing contributions in the sum, since
$R_{nl}(0)=0$ for all $l> 0$. When folded with a spherically symmetric test function, one
can reduce $\delta^{(3)}({\bf r})$ to ${1\over 4\pi} \,\delta(r)/r^2$.}:
$\sum_{m=1}^{\infty} R_{m S}(r) R_{mS}(0) = \delta(r)/r^2$,
together with the identity $R'_{nP}(0)=\lim_{r\to 0} R_{nP}(r)/r$, one can prove $\sum_m R_{mS}(0) {\cal Z}_{nm} = R'_{nP}(0)$.
Consequently, the UV divergent pieces in \eqref{chicJ:ee:HF:contribution:key:formula} reduce to
\beq
\widetilde{\cal A}^{(J)}_{\rm HF}
 = i \frac{e_c^2 \alpha^2\sqrt{N_c}}{ m_c^4}\sqrt{\frac{3}{4\pi}}
 R_{nP}'(0)
\left[{1\over \hat\epsilon_{\rm UV}}+2\ln{\mu\over m_c}+2 + \cdots \right].
\label{chicJ:ee:HF:contribution:UV:div}
\eeq
Just like what happens in Lamb shift, once adding the contributions from the leading Fock states,
\eqref{Leading:Fock:state:hard:contr:to:amplitude:chicJ:ee}, and those from the higher Fock states, \eqref{chicJ:ee:HF:contribution:UV:div},
the poles together with the affiliated $\ln\mu$ terms exactly cancel out, as promised~\footnote{One may invoke the KLN theorem to
justify why the IR divergences can be removed after including the higher Fock state contribution,
since both the higher and leading Fock components can be viewed as
{\it degenerate} for $\chi_{cJ}$.}.

\section{Full amplitudes and Bethe logarithms}

The finite, $\mu$-independent expressions of the leptonic decay amplitudes read
\begin{subequations}
\begin{align}
\widetilde{\cal A} [\chi_{c1}(nP)\to e^+ e^- ] & = \frac{2e_c^2 \alpha^2 \sqrt{N_c}}{m_c^4}
 \sqrt{\frac{3}{4\pi}} \left\{R'_{nP}(0) \left(1+\ln \frac{m_c}{2 \Delta E_{nP}}\right) + i\pi \sum_{m \leq n} R_{mS}(0) {\cal Z}_{nm} \right\}.
\label{reduced:amplitude:final:chic1:nP}
 \\
\widetilde{\cal A} [\chi_{c2}(nP)\to e^+ e^- ] & =  \frac{2e_c^2 \alpha^2 \sqrt{N_c}}{m_c^4}
\sqrt{\frac{3}{4\pi}} \left\{R_{nP}'(0)
\left({4\over 3}-{1\over 3}\ln 2 -{i\pi\over 3} + \ln {m_c \over 2\langle\Delta E_{nP}\rangle} \right)  +
i\pi \sum_{m \leq n} R_{mS}(0) {\cal Z}_{nm}  \right\},
\label{reduced:amplitude:final:chic2:nP}
\end{align}
\label{reduced:amplitude:final:chicJ}
\end{subequations}
which are complex-valued.
As is evident from \eqref{One:loop:integral:dim:reg},
the higher Fock component contribution also develops new imaginary parts whenever
it is kinetically permissible for the $E1$ transition $\chi_{cJ}(nP)\to \psi(mS)+\gamma$ to occur.

Analogous to the Bethe logarithm introduced in Lamb shift, here we have introduced the average binding energy
$\langle\Delta E_{nP}\rangle$ inherent to $\chi_{cJ}(nP)$ via
\begin{equation}
   \ln \langle\Delta E_{nP}\rangle  \equiv {1\over R_{nP}'(0)} \sum_{m=1}^\infty R_{mS}(0) {\cal Z}_{nm} \ln|E_{nP}-E_{mS}|.
\label{Def:Bethe:logarithm}
\end{equation}

Plugging the reduced amplitudes \eqref{reduced:amplitude:final:chicJ} into
\beq
\Gamma(\chi_{cJ}(nP) \to e^+e^-)= {1\over 2\pi} {1\over 2J+1}  M_{\chi_{cJ}}^4  \left| \widetilde{\cal A}^{(J)}\right|^2,
\label{From:amplitude:to:partial:widths}
\eeq
one then obtains the desired leptonic widths of $\chi_{cJ}$.

It is worth mentioning that the higher Fock component contributions respect heavy quark spin symmetry (HQSS),
due to spin-independent trait of the static potential and the $E1$ transition. Had the leading Fock state contribution been temporarily discarded,
one would anticipates $\Gamma^{(1)}_{\rm HF}: \Gamma^{(2)}_{\rm HF} = \frac{1}{3} : \frac{1}{5}$ as demanded by HQSS.
Of course, the leading Fock state contribution necessarily violates the HQSS, which is dominated by the hard loop contribution ($k\sim m_c$).
Therefore, the extent of the HQSS breaking in the leptonic widths \eqref{From:amplitude:to:partial:widths} may be viewed as
a barometer to gauge the relative importance of the higher Fock component contribution.

\section{Numerical results}\label{section:pheno}

\begin{table}[h]
    \centering
    \begin{tabular}{|c|c|c|c|c|c|}
    \hline
         & Logarithmic~\cite{Eichten:1995ch} & Buchmuller-Tye~\cite{Buchmuller:1980su} & Screened~\cite{Li:2009zu} & Power-law~\cite{Eichten:1995ch} &  Cornell~\cite{Eichten:1978tg} \\
    \hline
       $\Delta E_{1P}$[GeV]  & 0.34 & 0.32 & 0.29 & 0.35 & 0.28 \\
    \hline
       $\Gamma(\chi_{c1}\to e^+ e^-)$[eV]  & 0.32 & 0.33 & 0.42 & 0.49 & 0.70 \\
    \hline
       $\Gamma(\chi_{c2}\to e^+ e^-)$[eV]  & 0.12 & 0.13 & 0.17 & 0.19 & 0.28 \\
    \hline
       $ \Delta E_{2P}$[GeV]  & 0.26 & 0.27 & 0.24 & 0.27 & 0.25 \\
    \hline
    $\Gamma (c\bar{c}(2 ^3P_1) \to e^+ e^-)$[eV]  &0.72 &1.11 & 1.07 &1.22& 2.40 \\
    \hline
    $\Gamma (c\bar{c}(2 ^3P_2) \to e^+ e^-)$[eV]  & 0.29 & 0.45 &  0.45 & 0.49 &  1.01 \\
    \hline
    \end{tabular}
\caption{Bethe logarithms and the leptonic widths of $\chi_{c1,2}(1P,2P)$
predicted from five different potential models.}
\label{Table:Bethe:log:partial:widths:5:potential:models}
\end{table}

In numerical analysis, we employ the Buchm\"{u}ller-Tye potential~\cite{Buchmuller:1980su}
to generate central predictions for the leptonic widths of $\chi_{c1,2}(1P,2P)$.
The energy levels and the wave functions of numerous excited charmonium states are determined precisely by numerically solve
Schr\"{o}dinger equation. To quantify theoretical uncertainty, we also adopt several other influential
quark potential models compiled in \cite{Eichten:1995ch}, as well as the screened potential model
which are tailored to describe higher charmonia states above the open charm threshold~\cite{Li:2009zu}.

Substituting $\alpha=1/137$, $m_c=1.5$ GeV as well as the measured masses of $\chi_{cJ}(1P,2P)$~\cite{ParticleDataGroup:2022pth}
into \eqref{reduced:amplitude:final:chicJ} and \eqref{From:amplitude:to:partial:widths},
we then predict the leptonic decay rates of $\chi_{cJ}(nP)$ states with $n=1,2$.
Employing five popular quark potential models, we enumerate
in Table~\ref{Table:Bethe:log:partial:widths:5:potential:models}
our predictions of the Bethe logarithms and the leptonic widths for $\chi_{c1,2}(1P,2P)\to e^+e^-$.

The evaluation of the Bethe logarithms turns out to rapidly converge only after including the first five or six $\psi(mS)$ levels in
\eqref{Def:Bethe:logarithm}. As is evident in Table~\ref{Table:Bethe:log:partial:widths:5:potential:models},
the values of $\langle\Delta E_{nP} \rangle$ ranges from 200 to 300 MeV among different potential models, of the same order as $\Lambda_{\rm QCD}$.
This situation is in shark contrast with the Lamb shift in hydrogen atom,
where the counterpart $\Delta E_{2S}\approx 16.6$ Ry, much greater than the characteristic binding energy of hydrogen atom~\cite{Weinberg:QFT}.

\begin{table}[h]
\centering
\begin{tabular}{|c|c|c|c|c|c|c|}
\hline
       eV   & Exp. & This  work & Ref.~\cite{Yang:2012gk} & Ref.~\cite{Denig:2014fha} & Ref.~\cite{Kivel:2015iea} & Ref.~ \cite{Czyz:2016xvc}\\
    \hline
       $\Gamma(\chi_{c1}(1P)\to e^+ e^-)$ & $0.12^{+0.13}_{-0.08}$~\cite{BESIII:2022mtl} & $0.33^{+0.37}_{-0.01}$ & 0.039 & 0.10  & 0.09 &  0.43 \\
    \hline
       $\Gamma(\chi_{c2}(1P)\to e^+ e^-)$ &- & $0.13^{+0.15}_{-0.01}$ & 0.027 & - & 0.07 &  4.25 \\
    \hline
     $\Gamma(\chi_{c1}(2P)\to e^+ e^-)$ & $<0.32$ (90\% C.L.)~\cite{BESIII:2022ner}* & $1.11^{+1.29}_{-0.38}$ & - & $>0.03*$  & - &  - \\
    \hline
    $\Gamma(\chi_{c2}(2P)\to e^+ e^-)$ &- & $0.45^{+0.56}_{-0.16}$ & - & - & - &  - \\
\hline
\end{tabular}
\caption{Comparison between predictions from various groups and the available data for $\chi_{c1,2}(1P,2P)\to e^+e^-$. Note the entries
with asterisk implies that the $\chi_{c1}(2P)$ state should be replaced by $X(3872)$.
Note that the theoretical uncertainty inherent to charm quark mass is not included here, which would be potentially huge.}
\label{Predicted:partial:widths:Different:groups:vs:experiment}
\end{table}

In Table~\ref{Predicted:partial:widths:Different:groups:vs:experiment} we juxtapose our
predictions for $\chi_{c1,2}(1P,2P)\to e^-e^+$ with the predictions made by other groups,
including those obtained from the vector meson dominance model~\cite{Denig:2014fha,Czyz:2016xvc}.
Note our predicted values for $\Gamma[\chi_{c1,2}\to e^+e^-]$  are considerably larger than most of the preceding predictions.
It is interesting to note that our predicted leptonic width of $\chi_{c1}(1P)$
is compatible with the {\tt BESIII} measurement within $2\sigma$.
 We also note $\Gamma^{(1)}: \Gamma^{(2)}$ is not far from $5:3$, which implies that
the processes $\chi_{cJ}(nP)\to e^+e^-$ are dominated by contributions from the higher Fock components.

Since $\langle\Delta E_{nP} \rangle \sim \Lambda_{\rm QCD}$, one may anticipate that the original
calculation by Kuhn, Kaplan and Safiani~\cite{Kuhn:1979bb} would render reliable predictions by freezing the IR cutoff around the quarkonium binding energy.
Actually this is not the case. The non-logarithmic constants, especially the imaginary parts arising from the $\chi_{cJ}(nP)\to \psi(mS)+\gamma$ cuts
in \eqref{reduced:amplitude:final:chicJ} yield significant contributions.
A remarkable trait In Table~\ref{Predicted:partial:widths:Different:groups:vs:experiment} is that the leptonic widths of
$\chi_{cJ}(2P)$ are about three times bigger than those of the lowest-lying $\chi_{cJ}$, largely attributed to the inclusion of more
intermediate $\psi(mS)$ states in the sum.

It is particularly interesting to assess the observation prospect for $e^+e^-\to X(3872)$.  The early assignment of the $X(3872)$ as the
pure $P$-wave charmonium~\cite{Barnes:2003vb, Barnes:2005pb} or pure $D^0-\overline{D}^{0*}$ molecular state~\cite{Swanson:2003tb, Tornqvist:1993ng, Tornqvist:2004qy}
seems in conflict with the experimental data. In 2005 Suzuki suggested that $X(3872)$ is an admixture
between $c\bar{c}(2{}^3P_1)$ and the $D^0-\overline{D}^{0*}$ molecule~\cite{Suzuki:2005ha}.
Early investigation suggests that the $X(3872)$ is dominantly made of the $c\bar{c}(2{}^3P_1)$ component~\cite{Suzuki:2005ha, Li:2009zu}.
The following phenomenological studies~\cite{Ortega:2009hj,Coito:2012vf,Takizawa:2012hy,Butenschoen:2013pxa, Butenschoen:2019npa}
as well as the latest lattice analysis~\cite{Li:2024pfg}, favors a rather small $c\bar{c}(2{}^3P_1)$ content, say,
the probability to find the $c\bar{c}$ component inside $X(3872)$ is less than 10\%.
Within this picture, the large predicted leptonic width of $\chi_{c1}(2P)$, $1.11^{+1.29}_{-0.38}$ eV,
is not at odds with the  upper bound $0.32$ eV recently set by {\tt BESIII}~\cite{BESIII:2022ner}.
On the other hand, the future observation or nonobservation of $e^+e^-\to X(3872)$ would unambiguously decipher the $c\bar{c}$ content inside the $X(3872)$.

The $X(3930)$ is commonly assigned to be $\chi_{c2}(2P)$ (The $c\bar{c}$ component is estimated to make up $42\%$ fraction of the $X(3930)$~\cite{Ortega:2017qmg}).
The predicted leptonic width of $\chi_{c2}(2P)$ appears large enough to warrant the capability to observing the exclusive reaction $e^+e^-\to X(3930)$.
It is certainly rewarding for {\tt BESIII} experiment to search all the $e^+e^-\to \chi_{c2}, X(3872), X(3930)$ channels
with the aid of energy scan technique.


\section{Summary}\label{section:summary}

About four decades ago, Kuhn, Kaplan and Safiani discovered that the rare decay processes $\chi_{cJ}\to e^+e^-$ suffer from IR
divergence pathology, which implies the breakdown of NRQCD factorization for these hard exclusive reactions.
In this work we provide a systematic and satisfactory solution to this IR divergence problem.
It is interesting to recognize that the root and the cure of this IR divergence symptom is almost identical to
the Lamb shift in hydrogen atom. The key is to include the contribution of the higher Fock components of
the $\chi_{cJ}$ state, {\it viz.}, the $c\bar{c}(^3S_1)$ pair
accompanied with a long wavelength photon. After adding this new piece of contribution,
we then obtain the IR finite and factorization scale independent predictions for the partial widths of $\chi_{cJ}(nP)\to e^+e^-$.
Interestingly, the Bethe logarithms in this case has drastically different pattern relative to those affiliated with Lamb shift.
While the bulk of the Lamb shift of hydrogen is dominated by the continuum state contribution,
the bulk of the $\chi_{cJ}(nP)\to e^+e^-$ amplitudes not only comes from Bethe logarithm, but also stem from
the nonlogarithmic constants and absorptive part arising from the $\psi(mS)\gamma$ cuts.

We have used several different influential potential model to conduct numerical predictions.
The predicted partial width of $\chi_{c1}\to e^+e^-$ is reasonable agreement with the recent {\tt BESIII} measurement.
Our predictions for the leptonic widths of $\chi_{cJ}(1P,2P)$ are considerably greater than most of the preceding predictions in literature.
The leptonic width of $\chi_{c1}(2P)$ is particularly large, and we suggest that future observation of $e^+e^-\to X(3972)$
would unambiguously pin down the fraction of the $c\bar{c}$ content in $X(3872)$.
We are eagerly awaiting the {\tt BESIII} experiment to make a dedicated search for
all the $e^+e^-\to \chi_{c2}, X(3872), X(3930)$ signals,
so that our predictions can be critically tested in the near future.

\begin{acknowledgments}
We thank Yuping Guo, Zhewen Mo and Deshan Yang for useful discussions.
This work is supported in part by the National Natural Science Foundation of China
under Grants No.~11925506.
\end{acknowledgments}

\appendix

\section{Alternative derivations of higher Fock state contribution from EFT}
\label{Rederivation:from:pNRQCD+SCET}

The purpose of this appendix is to show that the key formulae in the main text can also be reproduced in the context of the modern EFT.
Since the physical system we are interested in consists the slowly-moving $c$ and $\bar{c}$,
the ultrasoft photon, and energetic electron and positron moving back-to-back,
the proper EFT to deduce the higher Fock component contribution to $\chi_{cJ}\to e^+e^-$
is the potential NRQCD/NRQED~\cite{Pineda:1997bj, Brambilla:1999qa, Brambilla:1999xf} in combination with SCET~\cite{Bauer:2000yr, Bauer:2001yt,Beneke:2002ph}:
\beq
\mathcal{L}_{\text{EFT}}=\int d^3 r\text{Tr}\left[S^{\dagger}\left(i\partial_{0}+\frac{\boldsymbol{\nabla}_r^2}{m_c}-V_S(r) \right) S\right] +
 {\cal L}_{\rm Maxwell}+
 V_{\text{E1}} + {\cal L}_{\text{SCET}_{\text I}},
\label{EFT:higher:Fock:state:contri}
\eeq
where $S(t,{\bf R},{\bf r})$ denotes a nonlocal interpolating current composed of nonrelativistic $c$ and $\bar{c}$ fields in the color-singlet channel, which is a $2\times 2$
matrix in the spinor space. For simplicity we set the center-of-mass coordinate $\boldsymbol{R}$ to the origin in the pNRQCD lagrangian.
Following the notation in \cite{Brambilla:2005zw}, we can express the propagator of the $S$ field projected onto the ${}^3S_1^{(1)}$ subspace
as
\beq
\left\langle 0\left| T\left[S_{ij}(t_1, {\bf r}_1) S^\dagger_{qt}(t_2, {\bf r}_2) \right] \right|0\right\rangle \Big|_{^3S_1^{(1)}}
=\int\!\! {d p^0\over 2\pi} e^{-i(t_1-t_2) p^0} \sum_n {i\over p^0-E_{nS}+i\epsilon}
{(\sigma^k)_{ij} R_{nS}(r_1)\over \sqrt{8\pi}} {(\sigma^k)_{qt} R_{nS}(r_2)\over \sqrt{8\pi}},
\label{S:field:propagator}
\eeq
where $R_{nS}(r)$ represents the radial wave function for the $\psi(nS)$ state.

$V_{\text{E1}}$ in \eqref{EFT:higher:Fock:state:contri} signifies the leading operator that mediates the
E1 transition,
\beq
V_{\text{E1}}= e e_c \int d^3 r \text{Tr} \left[ S^\dagger {\mathbf{A}_{em}\cdot \boldsymbol{\nabla}\over m_c} S\right].
\label{E1:transition:operator}
\eeq
Our pNRQCD/pNRQED treatment of the higher Fock component contributions is somewhat similar to
using pNRQED to account for the Lamb shift~\cite{Pineda:1997ie}.

The SCET lagrangian in \eqref{EFT:higher:Fock:state:contri} accounting for the collinear electron and the anti-collinear positron
interacting with the ultrasoft photon reads~\cite{Becher:2014oda}
\beq
{\cal L}_{\rm SCET_{\rm I}} =  \sum_{a=1,2} \bar{\xi}_{a}^{(0)}\, \frac{\slashed{\bar{n}}_a}{2}\, i n_a\cdot \partial \, \xi_{a}^{(0)}£¬
\eeq
where the decoupling transformation $\xi_a=Y_a\xi_{a}^{(0)}$ is made to eliminate
the explicit interaction between collinear lepton and ultrasoft photon,  with
$Y_a=\text{exp}\left\{ -ie\int_{0}^{\infty} ds \,n_a\cdot A^{\rm em} (sn_a)\right\}$.
The lightlike reference vectors are defined as $n_2=\bar{n}_1$ and $n_1=\bar{n}_2$, and normalized as $n_1\cdot n_2=2$.

 As depicted in Fig.~\ref{Feyn:diagrams:chicJ:to:ee}(b), we need an effective operator to mediate the reaction
$c\bar{c}(^3S_1)\to \gamma^*\to e^+ e^-$.
Since such an operator violates the conservation of the numbers of $c$/$\bar{c}$ separately,
it can not be added to the pNRQCD lagrangian, instead must be taken as an external current.

\begin{figure}[hbt]
\begin{center}
\includegraphics[width=0.8\textwidth]{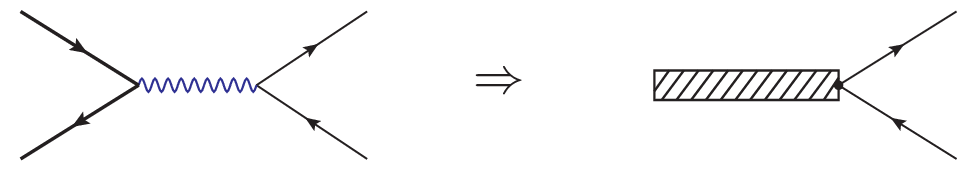}
\caption{Using the tree-level reaction $c\bar{c}(^3S_1)\to \gamma^*\to e^+ e^-$ to determine the effective operator $\mathcal{O}_{\rm eff}^{^3S_1 \to e^+ e^-}$.
The bar represents the color-singlet compound field $S(t,{\bf r}=0)$ in pNRQCD.
\label{Matching:onto:3S1:to:ee:operator}}
\end{center}
\end{figure}

The effective operator can be determined by enforcing the matrix element
$\langle e^+ e^- \vert \mathcal{O}_{\rm eff}^{^3S_1 \to e^+ e^-} \vert ^3S_1 \rangle$ to reproduce
the $c\bar{c}({}^3S^{(1)}_1)\to e^+e^-$ amplitude, as illustrated in Fig.~\ref{Matching:onto:3S1:to:ee:operator}.
It is legitimate to shrink the photon propagator to a point because the virtuality of the hard photon is about $4m_c^2$.
At lowest order in $\alpha_s$ and $v$, we obtain the following effective operator:
\beq
\mathcal{O}_{\rm eff}^{^3S_1 \to e^+ e^-}(0) =
-\frac{ie^2 e_c}{2m_c^2} \text{Tr} \left[ \sigma^{i} S(0) \right]\,\bar{\xi}_{n_2} Y_{n_2}^{\dagger} \gamma^{i} Y_{n_1} \xi_{n_1}(0).
\label{effective:operator:3S1:ee}
\eeq

We first prove that the contribution from the higher Fock state, \eqref{Higher:Fock:state:contr:to:S:matrix:chicJ:ee},
can be recast into the following time-ordered matrix element:
\bqa
{\cal A}^J_{\rm HF}     &= & \int_{-\infty}^\infty \!\! dt \, \left\langle e^+e^-| T\{  \mathcal{O}_{\rm eff}(0) \, iV_{E1}(t) \} |\chi_{cJ}(nP) \right\rangle
\nn\\
& = &  \int_{-\infty}^\infty \!\! dt \,  \int_{-\infty}^\infty \! {ds\over 2\pi} \, {i e^{ist}\over s+i \varepsilon}
\int \!\! {d^3 k\over (2\pi)^3 2 |{\bf k}|} \sum_{\rm Pol} \sum_{m=1}^\infty
\langle e^+e^-|\mathcal{O}_{\rm eff}(0)| \psi(mS)\gamma({\bf k}) \rangle \langle \psi(mS)\gamma({\bf k}) | iV_{E1}(t) |\chi_{cJ}(nP) \rangle
\nn\\
& = & \int \!\! {d^3 k\over (2\pi)^3 2 |{\bf k}|} \sum_{\rm Pol} \sum_{m=1}^\infty   \int_{-\infty}^\infty \!\! ds \,
\int_{-\infty}^\infty \!\! {dt\over 2\pi} \, {i e^{i(s+E_{mS}+|{\bf k}|-E_{\chi_{cJ}})t} \over s+i \epsilon}
\langle e^+e^-| \mathcal{O}_{\rm eff}(0) | \psi(mS)\gamma({\bf k}) \rangle \langle \psi(mS)\gamma({\bf k}) |iV_{E1}(0)  |\chi_{cJ}(nP) \rangle
\nn\\
& =&  \int \!\! {d^3 k\over (2\pi)^3 2 |{\bf k}|} \sum_{\rm Pol} \sum_{m=1}^\infty {i\over E_{\chi_{cJ}}-E_{mS}- |{\bf k}| +i \epsilon}
\langle e^+e^-| \mathcal{O}_{\rm eff}| \psi(mS)\gamma({\bf k}) \rangle \langle \psi(mS)\gamma({\bf k}) | iV_{E1} |\chi_{cJ}(nP) \rangle,
\label{HF:amplitude:time-ordered:product}
\eqa
where we have used the spectral representation for the Heaviside step function,
as well as inserted a complete set of intermediate states spanned by $\psi(mS)$ and photon.

With the knowledge of \eqref{S:field:propagator},
\eqref{E1:transition:operator} and \eqref{effective:operator:3S1:ee}, it is straightforward to work out
the matrix element of the time-ordered product \eqref{HF:amplitude:time-ordered:product},
with the aid of the familiar Lagrangian-based perturbation theory framed in the interaction picture.
Taking the $\chi_{c1}(nP)\to e^+e^-$ as a specific example (the extension to
$\chi_{c2}$ leptonic decay is trivial), and adopting the Coulomb gauge, we obtain
\begin{align}
{\cal A}^{(1)}_{\rm HF} & = \int dt \left\langle e^+e^-| T\{  \mathcal{O}_{\rm eff}^{^3S_1 \to e^+ e^-}(0)  \, iV_{E1}(t) \} |\chi_{c1}(nP) \right\rangle
\nn\\
& = { e^4 e_c^2 \sqrt{N_c} \over 2m_c^3}\sqrt{3\over 4\pi} \sum_{m=1}^\infty
R_{mS}(0) {\cal Z}_{nm}
\nn\\
& \times  \bar{u}(l_1) \gamma^{i}v(l_2)  \epsilon^{ijk} \varepsilon_{\chi_{c1}}^k  \, \mu^{2\epsilon} \!\!\int\! {d^D k \over (2 \pi)^D }
{1\over k^2+i\epsilon} {\delta^{jl}-\hat{k}^j\hat{k}^l \over E_{nP}-E_{mS}-k^0+i \epsilon} \left({n_1^l \over n_1 \cdot k} - {n_2^l \over n_2 \cdot k}\right),
\label{temp part2}
\end{align}
where dimensional regularization is invoked to regularize the UV divergence.
Recall the momenta carried by the electron/positron are $l^a= M_{\chi_c} n_a$ ($a=1,2$).
Integrating over $k^0$ using the method of residue, we then recover \eqref{chic1:ee:HF:contribution:key:formula},
the key formula in the main text obtained from the quantum mechanical perturbation theory.

\section{Leptonic width of $\chi_{bJ}(1P,2P)$}

\begin{table}[h]
    \centering
    \begin{tabular}{|c|c|c|c|c|c|}
    \hline
        & Logarithmic~\cite{Eichten:1995ch} & Buchmuller-Tye~\cite{Buchmuller:1980su} & Screened~\cite{Li:2009zu} & Power-law~\cite{Eichten:1995ch} &  Cornell~\cite{Eichten:1978tg}\\
    \hline
       $\Delta E_{1P}$[GeV]  & 0.34 & 0.30 & 0.37 & 0.33 & 0.29 \\
    \hline
       $\Gamma(\chi_{b1}\to e^+ e^-)$ ($\times10^{-3}$) [eV]  & 4.32 & 4.81 & 8.32 & 4.40 & 9.35 \\
    \hline
       $\Gamma(\chi_{b2}\to e^+ e^-)$ ($\times10^{-3}$) [eV]  & 2.03 & 2.24 & 3.63 & 2.10 & 4.20\\
    \hline
       $ \Delta E_{2P}$[GeV]  & 0.26 & 0.24 & 0.30 & 0.26 & 0.24 \\
    \hline
    $\Gamma (\chi_{b1}(2P) \to e^+ e^-)$ ($\times10^{-3}$) [eV]  & 5.34 & 6.77 & 10.69 & 5.79 & 12.69 \\
    \hline
    $\Gamma (\chi_{b2}(2P) \to e^+ e^-)$ ($\times10^{-3}$) [eV] & 2.60 & 3.25  & 4.80 & 2.85 &  6.07 \\
    \hline
    \end{tabular}
    \caption{Bethe logarithms and the partial widths of $\chi_{b1,2}(1P,2P)\to e^+e^-$ predicted from five different potential models.}
\label{Table:Bethe:log:partial:widths:5:potential:models:bottomonia}
\end{table}

For the sake of completeness, in this appendix we present the predictions for the leptonic widths
of $\chi_{bJ}(1P,2P)\to e^+ e^-$ and the affiliated $\langle \Delta E_{1P, 2P}\rangle $.
To estimate the theoretical uncertainties, we also choose five different potential models.
Substituting $\alpha=1/137$, $m_b=4.7$ GeV together with the measured masses of $\chi_{bJ}(1P,2P)$~\cite{ParticleDataGroup:2022pth}
into \eqref{reduced:amplitude:final:chicJ}, \eqref{Def:Bethe:logarithm} and \eqref{From:amplitude:to:partial:widths}, we
then obtain the desired results, which are enumerated in Table~\ref{Table:Bethe:log:partial:widths:5:potential:models:bottomonia}.

Curiously, the values of $\langle \Delta E_{1P, 2P}\rangle $ in bottomonium case
still range from 200 to 300 MeV, which is of order $\Lambda_{\rm QCD}$.
Unfortunately, the leptonic widths of $\chi_{bJ}$ are suppressed with respect to those of $\chi_{cJ}$
by three orders of magnitude, which makes it rather difficult to observe the $e^+e^-\to \chi_{bJ}(nP)$ channels
in the future {\tt Belle 2} experiments through energy scan technique.

\end{document}